\documentclass[preprint]{aastex}

\begin{document}

\title {Absolute Properties of the Eclipsing $\gamma$ Dor Star V404 Lyrae }
\author{Jae Woo Lee$^{1}$, Kyeongsoo Hong$^{2}$, Jae-Rim Koo$^{3}$, and Jang-Ho Park$^{1}$ }
\affil{$^1$Korea Astronomy and Space Science Institute, Daejeon 34055, Korea}
\affil{$^2$Institute for Astrophysics, Chungbuk National University, Cheongju 28644, Korea}
\affil{$^3$Department of Astronomy and Space Science, Chungnam National University, Daejeon 34134, Korea}
\email{jwlee@kasi.re.kr}

\begin{abstract}
We present the first high-resolution spectra for the eclipsing binary V404 Lyr showing $\gamma$ Dor pulsations, which we use 
to study its absolute properties. By fitting models to the disentangling spectrum of the primary star, we found that it has 
an effective temperature of $T_{\rm eff,1}=7,330 \pm 150$ K and a rotational velocity of $v_1\sin$$i=148\pm18$ km s$^{-1}$. 
The simultaneous analysis of our double-lined radial velocities and the pulsation-subtracted ${\it Kepler}$ data gives us 
accurate stellar and system parameters of V404 Lyr. The masses, radii, and luminosities are $M_1$ = 2.17$\pm$0.06 M$_\odot$, 
$R_1$ = 1.91$\pm$0.02 R$_\odot$, and $L_1$ = 9.4$\pm$0.8 L$_\odot$ for the primary, and $M_2$ = 1.42$\pm$0.04 M$_\odot$, 
$R_2$ = 1.79$\pm$0.02 R$_\odot$, and $L_2$ = 2.9$\pm$0.2 L$_\odot$ for the secondary. The tertiary component orbiting 
the eclipsing pair has a mass of $M_{\rm 3b}$ = 0.71$\pm$0.15 $M_\odot$ in an orbit of $P_{\rm 3b}$ = 642$\pm$3 days, 
$e_{\rm 3b}$ = 0.21$\pm$0.04, and $a_{\rm 3b}$ = 509$\pm$2 R$_\odot$. The third light of $l_3=4.1\pm0.2\%$ could be partly 
attributable to the K-type circumbinary object. By applying a multiple frequency analysis to the eclipse-subtracted light residuals, 
we detected 45 frequencies with signal to noise amplitude ratios larger than 4.0. Identified as independent pulsation modes, 
seven frequencies ($f_1-f_6$, $f_9$), their new pulsation constants, and the location in the Hertzsprung-Russell diagram indicate 
that the pulsating primary is a $\gamma$ Dor-type variable star. 
\end{abstract}

{}

\section{INTRODUCTION}

Double-lined eclipsing binary stars allow accurate and direct determination of their fundamental parameters, such as the mass 
and radius of each component at the percent level. Further, the binary components can be assumed to be coeval and to share 
the same evolutionary history. These constraints are very useful for asteroseismical studies of pulsating stars in binaries. 
Various types of pulsating stars are known as members of binary systems (see the review by Murphy 2018). We placed V404 Lyr 
(KIC 3228863, Gaia DR2 2051369705923204224) \textbf{on our spectroscopic observation program} for eclipsing binaries (EBs) in 2016. 
This paper is the sixth contribution in determining the physical properties of pulsating EBs by time-series spectroscopy and 
photometry (Hong et al. 2015, 2017, 2019; Koo et al. 2016; Lee et al. 2018).

V404 Lyr is one of many A-F spectral type stars observed by the {\it Kepler} satellite. Lee et al. (2014) reviewed the history 
of \textbf{the binary star observations} and presented the most recent comprehensive study from detailed analyses of 
the ${\it Kepler}$ time-series data and all available eclipse timings. They reported that the EB system is a quadruple star 
exhibiting $\gamma$ Dor pulsations and that the primary component is responsible for the high-order low-degree gravity ($g$) modes. 
However, \textbf{their binary parameters were preliminary} because no spectroscopic observations were available at the time. Here, 
we present the results from analyzing our high-resolution spectra and \textbf{the ${\it Kepler}$ light curve supporting our binary 
star model} and pulsational characteristics for V404 Lyr, consisting of a $\gamma$ Dor-type pulsator and its lobe-filling companion. 
The rest of the paper is organized as follows. Sections 2 and 3 describe the observations and spectral analysis, respectively. 
In Section 4, we present the binary modeling and pulsation frequenceis. Finally, Section 5 summarizes and discusses our conclusions.

\section{OBSERVATIONS AND DATA REDUCTIONS}

\subsection{{\it KEPLER} PHOTOMETRY}

The {\it Kepler} satellite observed our program target V404 Lyr in long-cadence mode almost continuously from Q0 to Q17. 
\textbf{This observing mode} has a sampling time of 29.42 min; 6.02-s exposures were co-added on board for 270 readouts. In their study, 
Lee et al. (2014) used the data from the {\it Kepler} EB catalogue (Pr\v sa et al. 2011; Slawson et al. 2011)\footnote{http://keplerebs.villanova.edu/}, 
where the Q12-13 and Q17 observations are not included. In this work, we used the simple aperture photometry (SAP) data 
retrieved from the MAST archive\footnote{http://archive.stsci.edu/kepler/}. The contamination level of these data is reported 
to be only 1 \% in the {\it Kepler} Input Catalog. The raw SAP data were detrended and normalized following the procedure described 
by Lee et al. (2017). \textbf{After the removal of obvious outliers filtered with 5$\sigma$ clipping by visual inspection}, 
the resultant {\it Kepler} observations comprise 65,901 data points, spanning $\sim$1470 days.

\subsection{GROUND-BASED SPECTROSCOPY}

High-resolution spectroscopy of V404 Lyr was carried out between 2016 April and 2018 April to measure the atmospheric parameters 
and radial velocities (RVs) of the component stars. The observations \textbf{were made with the fiber-fed echelle spectrograph BOES} 
(Kim et al. 2007) attached to the 1.8 m reflector at Bohyunsan Optical Astronomy Observatory (BOAO) in Korea. A total of 52 spectra 
were obtained using \textbf{the largest fiber, with a spectral resolution} of $R$ = 30,000. The exposure times of the pulsating EB 
were 1200 s to avoid orbital smearing, corresponding to 0.019 of the binary orbital period. \textbf{In addition, bias, THL, and 
ThAr images} were obtained before and after the observations. The BOES spectra were processed with the IRAF CCDPROC and ECHELLE package. 
Standard procedures were used for de-biasing, flat fielding, extraction of one-dimensional spectra, and wavelength calibration 
(Hong et al. 2015). The signal-to-noise (S/N) ratios were about 25 between 4000 $\rm \AA$ and 5000 $\rm \AA$.

\section{SPECTRAL ANALYSIS}

As in our previous papers (Hong et al. 2015; Lee et al. 2018), we chose isolated absorption lines of Fe I $\lambda$ 4957.61 
for the RV measurements of both components. The Fe I lines were fitted about 10 times by using two Gaussian functions 
with the $splot$ routine and deblending option in IRAF. The upper panel of Figure 1 shows the trailed spectra of V404 Lyr in 
the spectral region, and the lower panel presents the observed spectrum and its Gaussian fitting at an orbital phase of 0.74. 
The average RVs and the standard deviations ($\sigma$) of the primary and secondary stars are listed in Table 1 and displayed 
in Figure 2. 

The measured RVs of each star were applied to a sine curve of $V(t) = \gamma + K \sin[2 \pi (t-T_{\rm 0})/ P_{\rm orb}]$ to obtain 
the orbital parameters of V404 Lyr. Here, $\gamma$ is the systemic velocity, $K$ is the velocity semi-amplitude, $t$ is 
the observation time of each RV, and $T_{\rm 0}$ is the reference epoch of the orbital ephemeris. The orbital period was held fixed 
to be $P_{\rm orb}$ = 0.73094326 days obtained with the quadratic {\it plus} two-LTT (light-travel-time) ephemeris in the paper of 
Lee et al. (2014). We used the Levenberg-Marquart technique (Press et al. 1992) to evaluate the unknown parameters of $\gamma$, $K$, 
and $T_{\rm 0}$. The circular orbital solutions ($e$ = 0) are given in Table 2, together with related quantities, and are plotted in 
Figures 1 and 2. The rms residuals from each fit were calculated to be 11.7 km s$^{-1}$ and 18.1 km s$^{-1}$, respectively, 
for the primary and secondary components.

\textbf{To obtain the atmospheric parameters of our target star,} we extracted the separate spectra for each component using 
the spectral disentangling code FDB\textsc{inary} (Iliji\'c et al. 2004)\footnote{\url{http://sail.zpf.fer.hr/fdbinary/}}. 
Unfortunately, \textbf{the disentangling spectrum of the faint secondary has a low S/N ratio, so it is difficult to measure 
its atmospheric parameters reliably}. Generally, $\gamma$ Dor pulsators are main sequence or subgiant stars with spectral types 
of A7$-$F5 (Kaye et al. 1999), which correspond to temperatures ranging from 7800 K to 6500 K (Pecaut \& Mamajek 2013). 
For the effective temperature ($T_{\rm eff,1}$) and projected rotational velocity ($v_1\sin$$i$) of the primary component, 
we selected four absorption lines (Ca I $\lambda$4226, Fe I $\lambda$4271, Fe I $\lambda$4383, and Mg II $\lambda$4481) that 
are useful in the temperature classification of dwarf A$-$F stars according to the \textit{Digital Spectral Classification Atlas} 
given by R. O. Gray. We followed a procedure almost identical to that applied by Hong et al. (2017) and Lee et al. (2018).

First of all, the surface gravity was set to be log $g_1$ = 4.2 \textbf{based on the binary star modeling discussed in the next section}. 
\textbf{A microturbulent velocity of 2.0 km s$^{-1}$ and a solar metallicity of $[$Fe/H$]$ = 0} were assumed. Second, 
we constructed about 38,000 \textbf{synthetic spectra with ranges of} 6,000 $\le$ $T_{\rm eff,1}$ $\le$ 8,500 K (in steps of 10 K) 
and 50 $\le$ $v_1\sin$$i$ $\le$ 200 km s$^{-1}$ (in steps of 1 km s$^{-1}$) from the ATLAS9 atmosphere models (Kurucz 1993; 
Castelli \& Kurucz 2003). Third, the atmospheric parameters of the pulsating primary were found from a grid search minimizing 
the $\chi^2$ values between the synthetic models and the disentangling spectrum. Figure 3 shows the resulting $\chi^2$ diagrams 
for $T_{\rm eff,1}$ and $v_1\sin$$i$. We \textbf{obtained best fitting parameters} of $T_{\rm eff,1}$ = 7,330$\pm$150 K and 
$v_1\sin$$i$ = 148$\pm$18 km s$^{-1}$. The $T_{\rm eff,1}$ value is about 770 K hotter than the effective temperature (6,561 K) 
in the {\it Kepler} Input Catalogue (KIC; Kepler Mission Team 2009), \textbf{which is a much larger difference than the average 
difference} between the KIC and spectroscopic temperatures for dwarfs (Pinsonneault et al. 2012). \textbf{We examined how changes 
in metallicity ($\pm$0.1 in $[$Fe/H$]$) can affect our derived quantities (Katz et al. 1998; Torres et al. 2017). The result 
indicates that they have a minimal effect on $T_{\rm eff,1}$ and $v_1\sin$$i$.} Figure 4 presents \textbf{four selected regions} 
from the disentangling spectrum of the primary star, together with the synthetic spectra of 7180 K, 7330 K, and 7480 K.

\section{BINARY MODELING AND PULSATION FREQUENCIES}

The mass ratio ($q$) is one of the most important parameters in \textbf{binary modeling, because} many quantities are sensitive 
to the $q$ value. Lee et al. (2014) analyzed the {\it Kepler} light curve of V404 Lyr with the photometric $q$-search procedure. 
Their result indicated that the eclipsing pair is in a semi-detached configuration with a mass ratio of $q_{\rm ph}$ = 0.382. 
However, the photometric $q_{\rm ph}$ is very far from the spectroscopic mass ratio of $q_{\rm sp}$ = 0.660$\pm$0.015 presented 
in the previous section. This discrepancy may originate from partial eclipses and the possible existence of a third light source 
(Pribulla et al. 2003; Terrell \& Wilson 2005). To derive a unique set of binary parameters, we simultaneously modeled 
our RVs and the {\it Kepler} light curve of V404 Lyr by considering third-body and starspot effects. The 2007 version of 
the Wilson-Devinney synthesis code (Wilson \& Devinney 1971, van Hamme \& Wilson 2007; hereafter W-D) was used. 

The binary modeling of V404 Lyr was conducted in a method analogous to that for the double-lined eclipsing binaries OO Dra 
(Lee et al. 2018) and KIC 6206751 (Lee \& Park 2018). The effective temperature of the pulsating primary star was fixed at 
7,330$\pm$150 K \textbf{as measured by our spectral analysis in section 3}. The logarithmic limb-darkening coefficients were 
calculated from the tables of van Hamme (1993). The bolometric albedos ($A$) and the gravity-darkening exponents ($g$) were 
assumed to be $A_1$ = 1.0 and $g_1$ = 1.0 for the primary component, and $A_2$=0.5 and $g_2$ = 0.32 for its companion. 
\textbf{Also, synchronous rotations} for both components ($F_1=F_2=1.0$) were considered, and a circular orbit was adopted 
after several trials. The differential correction program of the W-D code was run until the correction of each parameter became 
smaller than its standard deviation. The result from this synthesis is given as Model 1 in the second to third columns of Table 3. 
The synthetic RV curves are displayed as red solid curves in Figure 2, and the synthetic light curves are presented as a blue curve 
in the top panel of Figure 5, where the corresponding residuals are plotted in the middle panel. The parenthesized errors in Table 3 
are the 1$\sigma$-values of the parameters obtained from the {\it Kepler} data in each quarter. 

Although some previous studies (e.g., Lee et al. 2017; Lee \& Park 2018) have shown that the light-curve parameters are immune 
to the light variations due to the pulsations, we cannot exclude this possibility for V404 Lyr. On the contrary, the pulsation 
frequencies could be \textbf{affected by changes in the derived binary parameters}. As the first step for frequency analyses, we split 
the {\it Kepler} data into intervals of 10 orbital periods. A total of 195 light curves were individually analyzed by simply 
adjusting both the ephemeris epoch ($T_0$) and the spot parameters except the colatitude in Model 1 of Table 3. The results 
are presented in Table 4, and the corresponding residuals are displayed as magnitude versus BJD in Figure 6. Then, 
the PERIOD04 program (Lenz \& Breger 2005) was applied to the outside-eclipse light residuals from the whole datasets. 
According to the successive prewhitening procedure for each frequency peak (Lee et al. 2014), we detected at least 45 frequencies 
with S/N amplitude ratios larger than 4.0 (Breger et al. 1993), which are listed in Table 5 and shown in Figure 7. The synthetic curve 
computed from the 45-frequency fit is displayed in the lower panel of Figure 6.  

We removed the pulsation signatures from the observed {\it Kepler} data. The pulsation-subtracted light curve was solved by using 
the Model 1 parameters as the initial values. The final results are listed as Model 2 in the fourth to fifth columns of Table 3. 
The binary parameters from both datasets (Model 1 and Model 2) are a good match to each other within their errors. This indicates 
that the light curve parameters of V404 Lyr are almost unaffected by multiperiodic oscillations. The pulsation-subtracted data and 
synthetic light curve are plotted as black circles and a red line, respectively, in the top panels of Figure 5. The light curve 
residuals from Model 2 are displayed in the bottom panel of this figure. 

Our binary modeling indicates that the eclipsing pair of V404 Lyr is a semi-detached binary \textbf{with a Roche-lobe filling secondary 
component}, and it allowed us to compute the absolute dimensions for each star. The results are listed in Table 6, together with 
those of Lee et al. (2014) for comparison. We determined the masses and radii of the EB system with an accuracy of about 2.6 \% 
and 1.2 \%, respectively. The distance to the system was calculated to be 671$\pm$41 pc, which is in satisfactory accord with 
the $Gaia$ distance of 634$\pm$14 pc taken from $Gaia$ DR2\footnote{https://gea.esac.esa.int/archive/} (1.578$\pm$0.033 mas; Gaia Collaboration et al. 2018). 
The fundamental parameters of V404 Lyr presented in this paper are quite different from those of Lee et al. (2014), 
which were \textbf{obtained from photometric data alone. We prefer our present results, which include} the double-lined RVs and 
atmospheric parameters from our high-resolution spectra.

\section{SUMMARY AND CONCLUSIONS}

In this article, \textbf{we report the first spectroscopic observations} of the eclipsing $\gamma$ Dor star V404 Lyr. From these spectra, 
\textbf{double-lined RVs were measured. The effective temperature} and projected rotational velocity of the pulsating primary were 
determined to be $T_{\rm eff,1}$ = 7,330$\pm$150 K and $v_1\sin$$i$ = 148$\pm$18 km s$^{-1}$, respectively. We analyzed the observed 
and pulsation-subtracted {\it Kepler} data, respectively, with our RV measurements. As seen in Table 3, the pulsation frequencies 
do not affect the binary parameters of the eclipsing system. The light and RV solutions indicate that V404 Lyr is 
\textbf{a classical Algol system, with a mass ratio} of $q$ = 0.653, a semi-major axis of $a$ = 5.23 R$_\odot$, an inclination angle 
of $i$ = 78$^\circ$.5, a temperature difference of $\Delta$($T_{1}$--$T_{2}$) = 1,712 K, and a third light of $l_{3}$ = 4.1 \%. 
The detached primary star fills its inner Roche lobe by about 91\%. The masses, radii, and luminosities of each component are 
$M_1$ = 2.17 M$_\odot$, $M_2$ = 1.42 M$_\odot$, $R_1$ = 1.91 R$_\odot$, $R_2$ = 1.79 R$_\odot$, $L_1$ = 9.4 L$_\odot$, and 
$L_2$ = 2.9 L$_\odot$. \textbf{In the Hertzsprung-Russell diagram (cf. Lee et al. 2016), the primary component is located close to 
the blue edge of the $\gamma$ Dor instability strip inside the main-sequence band, while the lobe-filling secondary lies above this band.} 

The synchronized rotations for the primary and secondary components were computed to be $v_{\rm 1,sync}=132.2\pm1.6$ km s$^{-1}$ 
and $v_{\rm 2,sync}=124.0\pm1.5$ km s$^{-1}$, respectively. The $v_{\rm 1,sync}$ value suggests that the primary star with 
the rotational velocity of $v_1\sin$$i=148\pm18$ km s$^{-1}$ is super-synchronous, but its does not provide sufficient evidence 
because the difference between them is within margins of errors. \textbf{We examined the sensitivity of the physical parameters of 
the binary system to the axial-to-mean orbital rotation ratio $F_1$. We find no notable differences between synchronous rotation 
$F_1 = 1.0$ or asynchronous $F_1 = 1.1$. However, the close separation of the components of V404 Lyr lead us to believe the components 
are in synchronous rotation or nearly so.} 

The eclipse timing analyses of Lee et al. (2014) indicated that V404 Lyr is probably a quadruple star system, with LTT periods 
of $P_3$ = 649 days and $P_4$ = 2154 days and minimum masses of $M_3$ = 0.47 M$_\odot$ and $M_4$ = 0.047 M$_\odot$, where $M_1+M_2$ = 
1.87 M$_\odot$ \textbf{was assumed}. If we use the new value of $M_1+M_2$ = 3.59 M$_\odot$, these masses become $M_3$ = 0.71 M$_\odot$ 
and $M_4$ = 0.073 M$_\odot$, respectively. In this paper, \textbf{the improved physical parameters} of the third body were obtained 
by assuming its orbit to be coplanar with that of the eclipsing pair. As a result, the tertiary component has a period of 
$P_{\rm 3b}$ = 642 days, an eccentricity of $e_{\rm 3b}$ = 0.21, and a semi-major axis of $a_{\rm 3b}$ = 509 R$_\odot$, and its mass 
become $M_3$ = 0.71 M$_\odot$. The third-body mass corresponds to a spectral type of about K4.5V, and the bolometric luminosity is 
calculated to be $L_3$ = 0.21 $L_{\sun}$ (Pecaut \& Mamajek 2013), which contributes about 2 \% to the total light of this system. 
The third light of $l_3 \approx 4\%$ detected in this work may be attributed to the K-type circumbinary object gravitationally 
bound to V404 Lyr and/or the contamination due to the nearby star. \textbf{It is nearly impossible to find such a companion from 
our spectroscopic observations due to its faintness and low S/N ratio. Instead, the predicted semi-amplitude of the systemic RV change 
for the eclipsing pair by the circumbinary object is about 6 km s$^{-1}$, which may be detected with high-resolution spectroscopy 
extending over the course of a few years. On the other hand, the ($V-K_{\rm s}$) color index for such K-type stars is about $+$2.8 mag 
(Pecaut \& Mamajek 2013), so the third body can be as bright as $K_{\rm s} \sim$ 12 mag. However, because the maximum angular separation 
between the eclipsing pair and its companion is less than 4 mas, the circumbinary object cannot be revealed easily using interferometry. } 

The PERIOD04 periodogram for the eclipse-subtracted light residuals revealed 45 oscillation signals with S/N $>$ 4.0, which is five 
more than those detected by Lee et al. (2014). The main frequencies of $f_1-f_6$ and $f_9$ found in this paper are exactly the same 
as the $\gamma$ Dor pulsations identified in the previous paper. In contrast, most of the other frequencies may be orbital harmonics 
or combination frequencies. $\gamma$ Dor stars pulsate in high-order $g$ modes with multiple periods between 0.4 and 3 days, and 
they have the pulsation constants of $Q >$ 0.23 days (Henry et al. 2005). \textbf{To obtain the $Q$ values} for $f_1-f_6$ and $f_9$, 
we used the relation between the pulsation period ($P_{\rm pul}$) and the mean density ($\rho$) defined as 
$Q$ = $P_{\rm pul}$$\sqrt{\rho / \rho_\odot}$. From this equation and the $\rho _1$ value given in Table 6, we obtained 
$0.268<Q<0.307$ for the period range of $P_{\rm pul}=0.473-0.539$ days. The results confirm that the primary component is 
a $\gamma$ Dor-type pulsating star. The physical properties of V404 Lyr match well the possible correlations ($P_{\rm pul}-P_{\rm orb}$, 
$P_{\rm pul}-Q$, and $P_{\rm pul}-R/a$) for eclipsing $\gamma$ Dor stars presented by Ibanoglu et al. (2018). 
\textbf{Because the primary components in semi-detached EBs have evolved with the tidal force and mass accretion from their companions, 
the pulsating characteristics of our target star may be affected by the binary effects, unlike those of single pulsators. 
As in case of KIC 620751 (Lee \& Park 2018), the ratios of the orbital-to-pulsational frequencies for V404 Lyr are 
$f_{\rm orb}$:$f_{\rm 1-6,9}$ $\simeq$ 2:3. These indicate that the $\gamma$ Dor pulsations may be excited by the tidal interaction 
between the component stars. The pulsating EBs with reliable absolute properties can be applied to study the influence of tides on 
stellar structure.}

\acknowledgments{ }
The authors wish to thank the staff of BOAO for assistance during our observations. \textbf{This paper includes data collected by 
the {\it Kepler} mission. Funding for the {\it Kepler} mission is provided by the NASA Science Mission directorate. Some of the data 
presented in this paper were obtained from the Mikulski Archive for Space Telescopes (MAST). We appreciate the careful reading and 
valuable comments of the anonymous referee.} This research has made use of the Simbad database maintained at CDS, Strasbourg, France, 
and was supported by the KASI grant 2019-1-830-03. K.H. was supported by the grant Nos. 2017R1A4A1015178 and 2019R1I1A1A01056776 of 
the National Research Foundation (NRF) of Korea, respectively.

\clearpage
\begin{figure}
\includegraphics[scale=0.9]{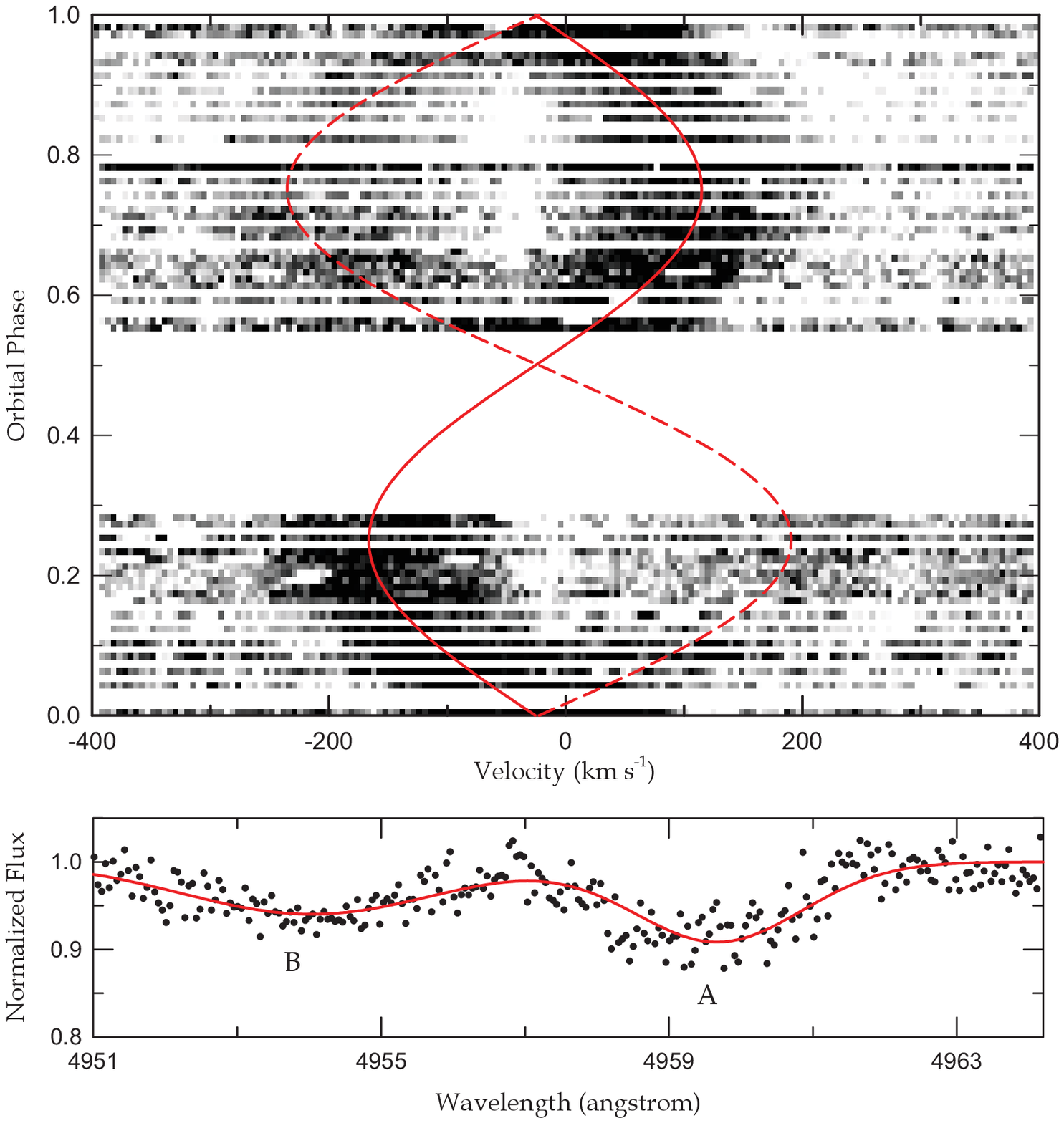}
\caption{The upper panel displays the trailed spectra of V404 Lyr in the Fe I $\lambda 4957.61$ region. The solid and 
dashed lines track the orbital motions of the primary (A) and secondary (B) components, respectively. In the lower panel, 
the circle and line represent the observed spectrum and its Gaussian fitting at an orbital phase of 0.74, respectively. }
\label{Fig1}
\end{figure}

\begin{figure}
\includegraphics[]{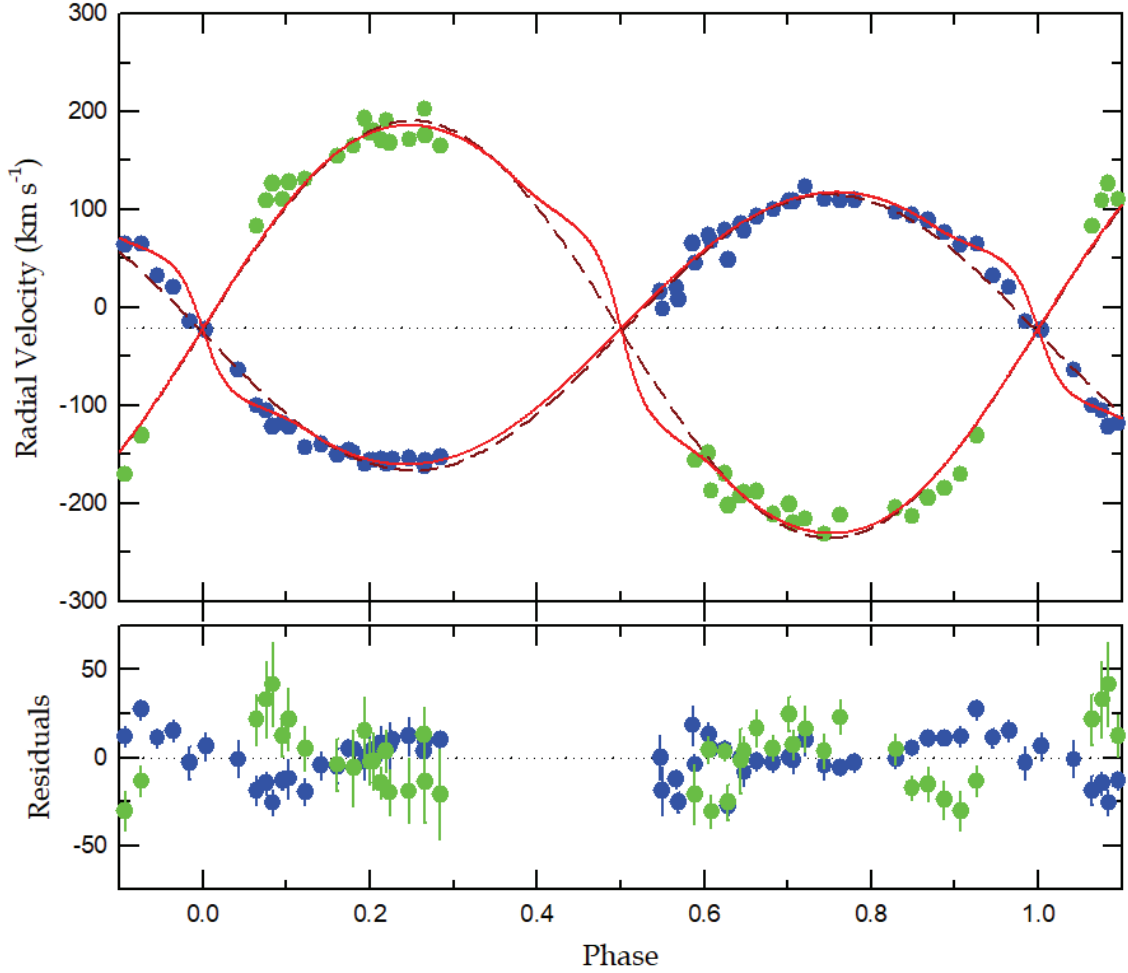}
\caption{RV curves of V404 Lyr with fitted models. The blue and green circles are the primary and secondary measurements, 
respectively. The dashed curves in the upper panel were obtained by separately fitting a sine wave to each RV curve, 
while the solid curves represent the results from a consistent light and RV curve analysis including proximity effects with 
the W-D code. The dotted line represents the system velocity of $-$21.4 km s$^{-1}$. The lower panel shows the residuals 
between observations and sine-curve fits, where each vertical line is an error bar for each RV measurement. 
\textbf{The non-sinusoidal RV residuals are artifacts of the solution and not physical.} }
\label{Fig2}
\end{figure}

\begin{figure}
\includegraphics[scale=0.9]{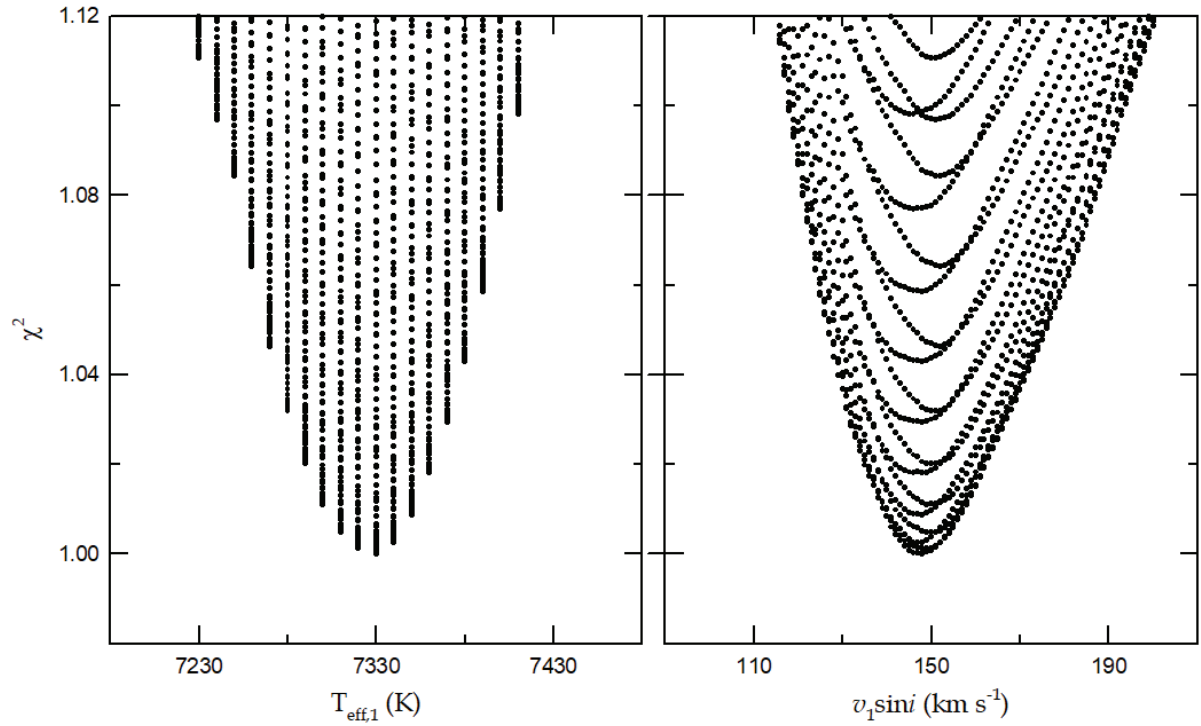}
\caption{$\chi^2$ diagrams of the effective temperature (left) and the projected rotational velocity (right) of the primary 
star. The surface gravity and metallicity are fixed as $\log$ $g$ = 4.2 and [Fe/H] = 0.0, respectively. }
\label{Fig3}
\end{figure}

\begin{figure}
\includegraphics[scale=0.8]{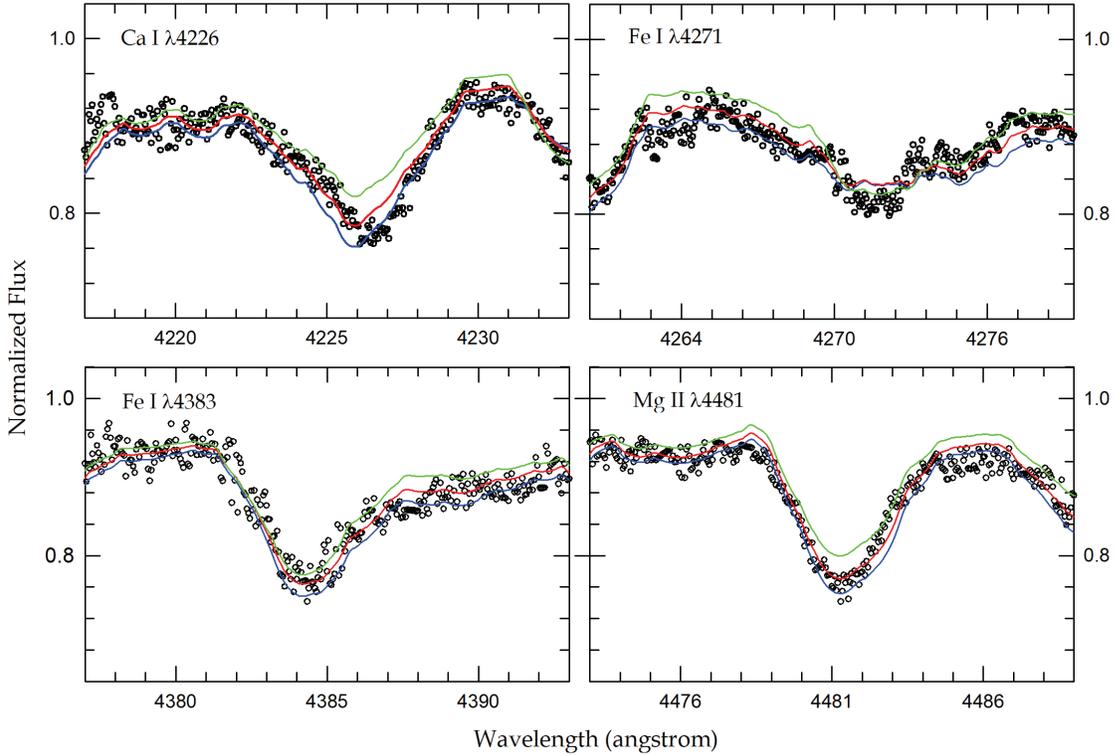}
\caption{Four spectral regions of the primary star. The open circles represent the disentangling spectrum obtained by 
the FDBinary code. The blue, red, and green lines represent the synthetic spectra of 7180 K, 7330 K, and 7480 K, respectively, 
interpolated from atmosphere models (Kurucz 1993; Castelli \& Kurucz 2003), where log $g_1 = 4.2$, [Fe/H] = 0.0, and 
$v_1\sin$$i$ = 148 km s$^{-1}$. }
\label{Fig4}
\end{figure}

\begin{figure}
\includegraphics[]{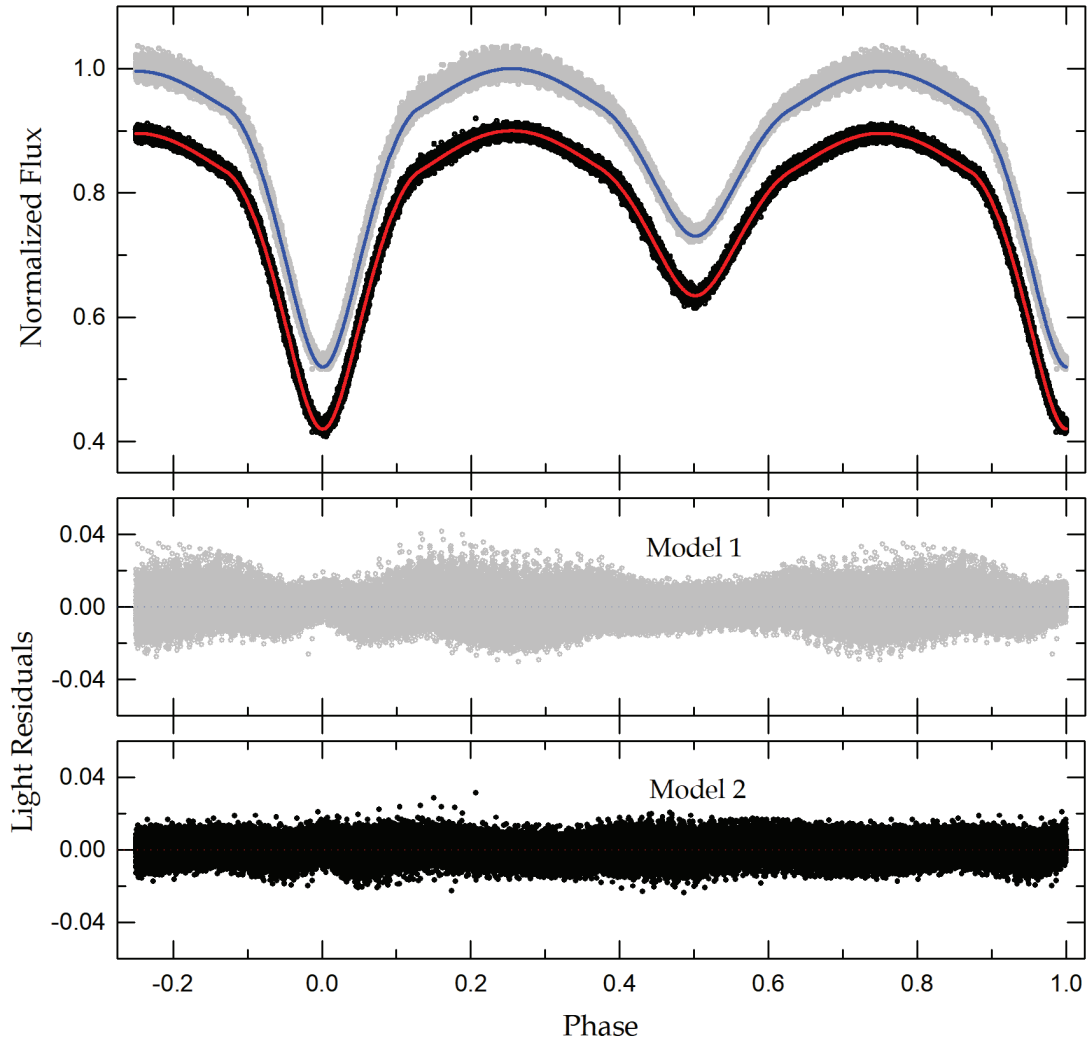}
\caption{Binary light curve of V404 Lyr before (gray circle) and after (black circle) subtraction of the pulsation signatures from 
the observed {\it Kepler} data. The blue and red solid curves were computed with the Model 1 and Model 2 parameters of Table 3, 
respectively. The corresponding residuals from Models 1 and 2 are plotted at the middle and bottom panels, respectively. }
\label{Fig5}
\end{figure}

\begin{figure}
\includegraphics[]{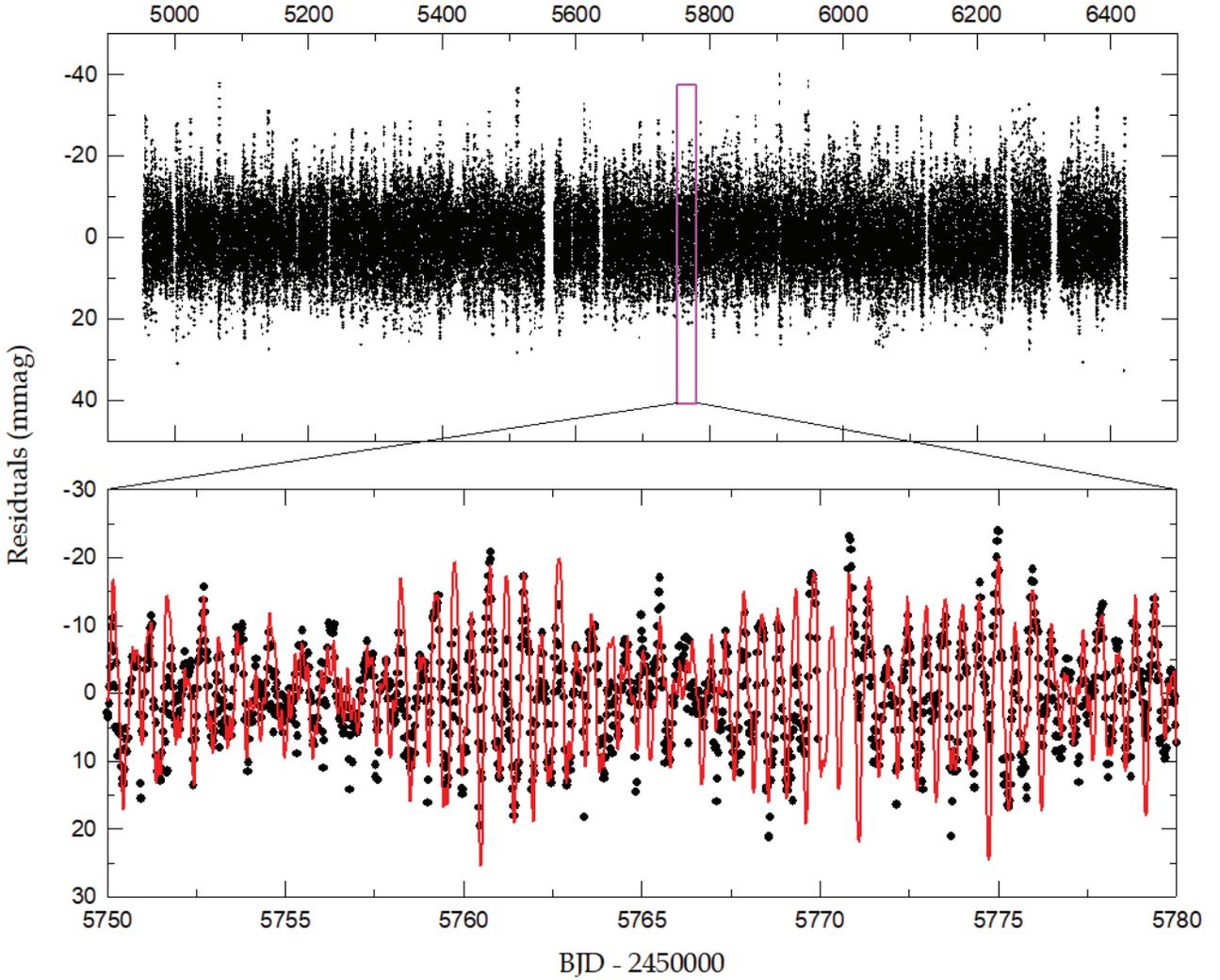}
\caption{Light curve residuals after subtraction of the binarity effects from each of 195 light curves combined at intervals of 
10 orbital periods in the observed {\it Kepler} data. The lower panel presents a short section of the residuals marked using 
the inset box of the upper panel. \textbf{The measurement errors of the points are smaller than 0.01 mmag, and their error bars 
cannot be seen in the plot.} The synthetic curve is computed from the 45-frequency fit to the outside-eclipse data. 
 }
\label{Fig6}
\end{figure}

\begin{figure}
\includegraphics[]{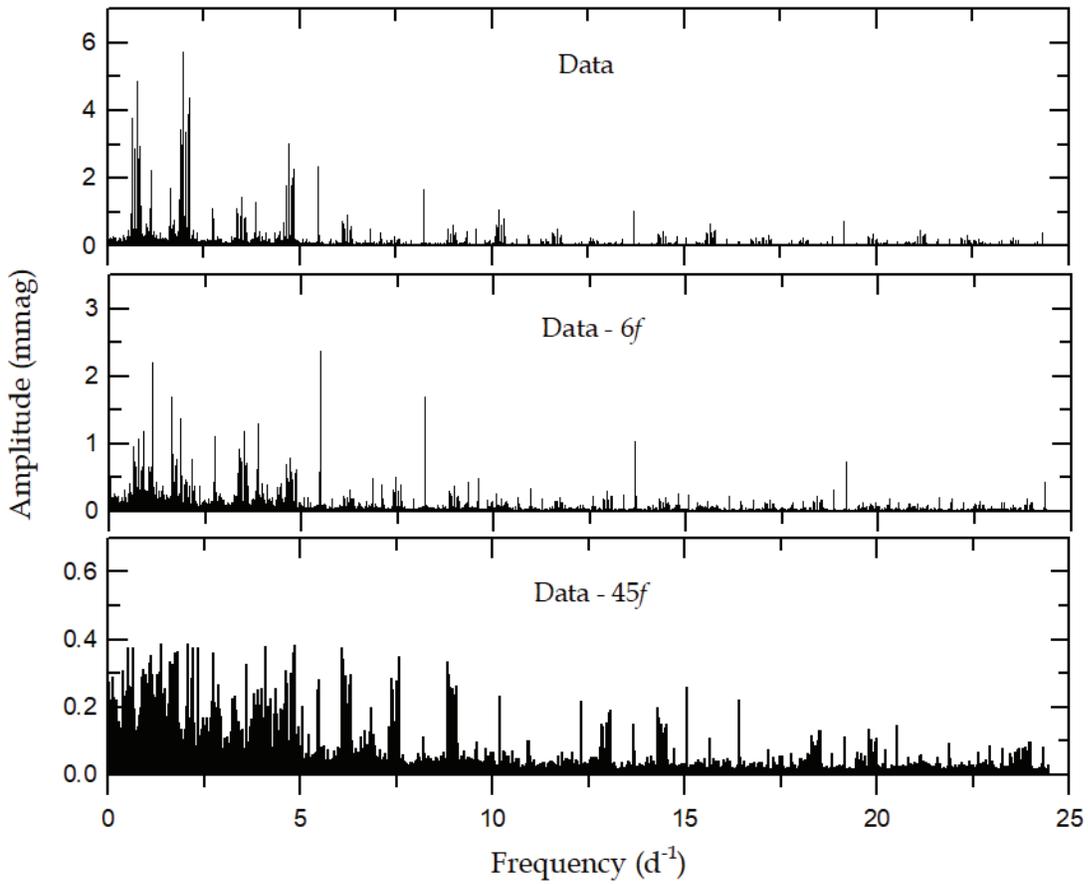}
\caption{Amplitude spectra before (top panel) and after prewhitening of the first 6 frequencies (middle panel) and all 45 frequencies 
(bottom panel) from the PERIOD04 program. The frequency analysis was applied to the entire outside-eclipse light residuals. }
\label{Fig7}
\end{figure}

\clearpage                                                                                                           
\begin{deluxetable}{lrrrr}
\tablewidth{0pt}                    
\tabletypesize{\small}   
\tablecaption{Radial Velocities of V404 Lyr }                                                                            
\tablehead{    
\colhead{BJD}          & \colhead{$V_{1}$}       & \colhead{$\sigma_1$}    &  \colhead{$V_{2}$}      &  \colhead{$\sigma_2$}    \\                                            
\colhead{(2,450,000+)} & \colhead{(km s$^{-1}$)} & \colhead{(km s$^{-1}$)} & \colhead{(km s$^{-1}$)} & \colhead{(km s$^{-1}$)}  
}                                                                                                 
\startdata
7,497.2713             &  $ -105.3 $             &   6.4                   &  $  109.0 $             &   21.4                   \\
7,497.2855             &  $ -118.6 $             &   5.0                   &  $  110.5 $             &   11.9                   \\
7,503.1907             &  $ -145.6 $             &   5.2                   &  $  \dots $             &   \dots                  \\
7,503.2049             &  $ -160.1 $             &  10.6                   &  $  192.7 $             &   19.6                   \\
7,503.2191             &  $ -154.6 $             &   9.8                   &  $  170.6 $             &    9.2                   \\
7,521.1223             &  $  108.1 $             &   7.2                   &  $ -220.5 $             &    8.4                   \\
7,522.2338             &  $ -154.7 $             &   4.9                   &  $  \dots $             &   \dots                  \\
7,522.2480             &  $ -154.2 $             &  11.0                   &  $  171.5 $             &   18.5                   \\
7,522.2622             &  $ -156.5 $             &   5.6                   &  $  175.8 $             &   23.2                   \\
7,522.2754             &  $ -153.1 $             &   4.9                   &  $  165.0 $             &   25.7                   \\
7,523.2006             &  $ -1.1   $             &  14.0                   &  $  \dots $             &   \dots                  \\
7,523.2148             &  $   8.4  $             &   5.9                   &  $  \dots $             &   \dots                  \\
7,523.2291             &  $  45.0  $             &   5.8                   &  $ -156.3 $             &   17.1                   \\
7,523.2433             &  $  67.6  $             &   4.0                   &  $ -187.5 $             &    9.5                   \\
7,523.2582             &  $  48.8  $             &   5.5                   &  $ -202.3 $             &   10.1                   \\
7,523.2721             &  $  78.6  $             &   7.6                   &  $ -189.5 $             &    8.1                   \\
7,901.0966             &  $  16.0  $             &  12.5                   &  $  \dots $             &   \dots                  \\
7,901.1107             &  $  19.6  $             &   5.5                   &  $  \dots $             &   \dots                  \\
7,901.1248             &  $  65.4  $             &  11.2                   &  $  \dots $             &   \dots                  \\
7,901.1389             &  $  74.0  $             &   6.8                   &  $ -149.2 $             &    7.8                   \\
7,901.1529             &  $  78.5  $             &   4.8                   &  $ -169.7 $             &    5.2                   \\
7,901.1670             &  $  85.1  $             &   6.4                   &  $ -192.1 $             &   18.5                   \\
7,901.1811             &  $  92.6  $             &   5.7                   &  $ -188.0 $             &   10.4                   \\
7,901.1952             &  $  99.6  $             &   5.5                   &  $ -211.5 $             &    7.2                   \\
7,901.2093             &  $ 108.6  $             &   4.6                   &  $ -201.2 $             &    9.6                   \\
7,901.2234             &  $ 123.1  $             &   9.6                   &  $ -215.8 $             &   13.2                   \\
8,018.9844             &  $  97.2  $             &   5.9                   &  $ -204.7 $             &    8.6                   \\
8,018.9985             &  $  94.2  $             &   4.6                   &  $ -213.2 $             &    6.8                   \\
8,019.0127             &  $  88.7  $             &   4.8                   &  $ -194.5 $             &    9.9                   \\
8,019.0269             &  $  76.6  $             &   4.4                   &  $ -184.8 $             &   11.1                   \\
8,019.0410             &  $  63.9  $             &   5.6                   &  $ -170.3 $             &   11.2                   \\
8,019.0552             &  $  64.4  $             &   5.7                   &  $ -131.0 $             &    8.7                   \\
8,019.0693             &  $  32.5  $             &   5.6                   &  $  \dots $             &   \dots                  \\
8,019.0835             &  $  20.2  $             &   5.9                   &  $  \dots $             &   \dots                  \\
8,019.0977             &  $ -14.9  $             &   9.4                   &  $  \dots $             &   \dots                  \\
8,019.1118             &  $ -22.8  $             &   8.1                   &  $  \dots $             &   \dots                  \\
8,019.1401             &  $ -63.9  $             &  10.6                   &  $  \dots $             &   \dots                  \\
8,043.0428             &  $  110.0 $             &   7.4                   &  $ -231.9 $             &    9.8                   \\
8,043.0570             &  $  108.8 $             &   4.6                   &  $ -211.9 $             &    9.7                   \\
8,043.0694             &  $  109.6 $             &   5.3                   &  $  \dots $             &   \dots                  \\
8,220.2678             &  $ -156.4 $             &   8.4                   &  $  179.7 $             &   16.2                   \\
8,220.2820             &  $ -158.0 $             &  13.7                   &  $  168.0 $             &   12.9                   \\
8,220.3123             &  $ -161.8 $             &   5.4                   &  $  202.5 $             &   16.1                   \\
8,228.2059             &  $ -100.0 $             &   8.2                   &  $   83.1 $             &   14.7                   \\
8,228.2200             &  $ -122.1 $             &   7.0                   &  $  126.3 $             &   23.9                   \\
8,228.2342             &  $ -122.6 $             &  11.2                   &  $  127.8 $             &   17.6                   \\
8,228.2483             &  $ -143.1 $             &   7.6                   &  $  130.8 $             &   12.6                   \\
8,228.2625             &  $ -139.7 $             &   8.2                   &  $  \dots $             &   \dots                  \\
8,228.2767             &  $ -150.1 $             &   9.6                   &  $  154.4 $             &   14.8                   \\
8,228.2908             &  $ -148.5 $             &  10.4                   &  $  164.8 $             &   21.5                   \\
8,228.3050             &  $ -156.4 $             &  10.0                   &  $  177.6 $             &   13.0                   \\
8,228.3192             &  $ -159.2 $             &  10.7                   &  $  190.5 $             &   11.9                   \\
\enddata                                                                                                             
\end{deluxetable}

\begin{deluxetable}{lcc}
\tablewidth{0pt}
\tablecaption{Orbital Elements of V404 Lyr Derived with Sine-Curve Fits }
\tablehead{
\colhead{Parameter}          & \colhead{Primary}         & \colhead{Secondary}
}                                                         
\startdata                                                
$T_0$ (BJD)                  & \multicolumn{2}{c}{2,458,019.1090$\pm$0.0021}         \\
$P_{\rm orb}$ (day)$\rm ^a$  & \multicolumn{2}{c}{0.73094326}                        \\
$\gamma$ (km s$^{-1}$)       & $-$25.7$\pm$1.7           & $-$22.4$\pm$3.1           \\
$K$ (km s$^{-1}$)            & 140.7$\pm$2.2             & 213.1$\pm$3.8             \\ 
$a\sin$$i$ ($R_\odot$)       & 2.032$\pm$0.032           & 3.078$\pm$0.055           \\
$M\sin ^3$$i$ ($M_\odot$)    & 2.020$\pm$0.062           & 1.334$\pm$0.039           \\
$q $ (= $M_2/M_1$)           & \multicolumn{2}{c}{0.660$\pm$0.015}                   \\
\enddata
\tablenotetext{a}{Fixed.}
\end{deluxetable}

\begin{deluxetable}{lccccc}
\tabletypesize{\small}  
\tablewidth{0pt} 
\tablecaption{Light and RV Parameters of V404 Lyr }
\tablehead{
\colhead{Parameter}                      & \multicolumn{2}{c}{Model 1$\rm ^a$}         && \multicolumn{2}{c}{Model 2$\rm ^b$}         \\ [1.0mm] \cline{2-3} \cline{5-6} \\[-2.0ex]
                                         & \colhead{Primary} & \colhead{Secondary}     && \colhead{Primary} & \colhead{Secondary}         
}
\startdata 
$T_0$ (BJD)                              & \multicolumn{2}{c}{2,454,953.53250(14)}     && \multicolumn{2}{c}{2,454,953.53254(14)}     \\
$P_{\rm orb}$ (day)                      & \multicolumn{2}{c}{0.730944022(54)}         && \multicolumn{2}{c}{0.730944147(54)}         \\
$a$ (R$_\odot$)                          & \multicolumn{2}{c}{5.270(60)}               && \multicolumn{2}{c}{5.225(60)}               \\
$\gamma$ (km s$^{-1}$)                   & \multicolumn{2}{c}{$-$21.4(1.3)}            && \multicolumn{2}{c}{$-$21.4(1.3)}            \\
$K_1$ (km s$^{-1}$)                      & \multicolumn{2}{c}{141.2(1.9)}              && \multicolumn{2}{c}{140.1(1.9)}              \\
$K_2$ (km s$^{-1}$)                      & \multicolumn{2}{c}{216.5(3.3)}              && \multicolumn{2}{c}{214.4(3.3)}              \\
$q$                                      & \multicolumn{2}{c}{0.652(13)}               && \multicolumn{2}{c}{0.653(13)}               \\
$i$ (deg)                                & \multicolumn{2}{c}{78.614(66)}              && \multicolumn{2}{c}{78.526(58)}              \\
$T$ (K)                                  & 7330(150)         & 5647(87)                && 7330(150)         & 5618(82)                \\
$\Omega$                                 & 3.5001(32)        & 3.1580                  && 3.4685(28)        & 3.1600                  \\
$X$, $Y$                                 & 0.648, 0.249      & 0.647, 0.203            && 0.648, 0.249      & 0.647, 0.203            \\
$x$, $y$                                 & 0.610, 0.284      & 0.696, 0.246            && 0.610, 0.284      & 0.696, 0.246            \\
$l$/($l_{1}$+$l_{2}$+$l_{3}$)            & 0.7107(18)        & 0.2489                  && 0.7187(16)        & 0.2404                  \\
$l_{3}$$\rm ^c$                          & \multicolumn{2}{c}{0.0404(23)}              && \multicolumn{2}{c}{0.0409(21)}              \\
$r$ (pole)                               & 0.3467(15)        & 0.3209(17)              && 0.3506(13)        & 0.3211(15)              \\
$r$ (point)                              & 0.3909(37)        & 0.4561(21)              && 0.3980(33)        & 0.4563(18)              \\
$r$ (side)                               & 0.3598(18)        & 0.3356(18)              && 0.3643(16)        & 0.3358(16)              \\
$r$ (back)                               & 0.3752(25)        & 0.3677(18)              && 0.3807(22)        & 0.3678(16)              \\
$r$ (volume)$\rm ^d$                     & 0.3611(22)        & 0.3428(18)              && 0.3657(19)        & 0.3430(16)              \\ [1.0mm]
\multicolumn{6}{l}{Third-body Parameters:}                                                                                            \\        
$a_{\rm 3b}$($R_\odot$)                  & \multicolumn{2}{c}{514.1(1.5)}              && \multicolumn{2}{c}{508.8(1.5)}              \\        
$i_{\rm 3b}$ (deg)                       & \multicolumn{2}{c}{78}                      && \multicolumn{2}{c}{78}                      \\        
$e_{\rm 3b}$                             & \multicolumn{2}{c}{0.232(38)}               && \multicolumn{2}{c}{0.209(39)}               \\        
$\omega_{\rm 3b}$  (deg)                 & \multicolumn{2}{c}{223(10)}                 && \multicolumn{2}{c}{216(10)}                 \\        
$P_{\rm 3b}$ (day)                       & \multicolumn{2}{c}{643.9(2.8)}              && \multicolumn{2}{c}{642.0(2.8)}              \\        
$T_{\rm c,3b}$ (BJD)                     & \multicolumn{2}{c}{2,454,830(11)}           && \multicolumn{2}{c}{2,454,828(11)}           \\        
$M_3$/($M_{1}$+$M_{2}$)                  & \multicolumn{2}{c}{0.196(41)}               && \multicolumn{2}{c}{0.197(41)}               \\ [1.0mm]
\multicolumn{6}{l}{Spot Parameters:}                                                                                                  \\        
Colatitude (deg)                         & \dots             & 57.29(62)               && \dots             & 72.8(1.5)               \\        
Longitude (deg)                          & \dots             & 340.5(1.2)              && \dots             & 337.5(2.5)              \\        
Radius (deg)                             & \dots             & 24.51(36)               && \dots             & 23.19(53)               \\        
$T$$\rm _{spot}$/$T$$\rm _{local}$       & \dots             & 0.9433(12)              && \dots             & 0.9618(11)              \\
$\Sigma W(O-C)^2$                        & \multicolumn{2}{c}{0.0075}                  && \multicolumn{2}{c}{0.0053}                  \\[1.0mm]
\enddata
\tablenotetext{a}{Result from the observed data.}
\tablenotetext{b}{Result from the pulsation-subtracted data.}
\tablenotetext{c}{Value at 0.25 phase. }
\tablenotetext{d}{Mean volume radius. }
\end{deluxetable}

\begin{deluxetable}{lccc}
\tablewidth{0pt}
\tablecaption{Epoch and Spot Parameters for 195 datasets at Intervals of 10 Orbital Periods  }
\tablehead{
\colhead{Epoch}    & \colhead{Longitude}  & \colhead{Angular Radius}   & \colhead{Temp. Factor}   \\
\colhead{(BJD)}    & \colhead{(deg)}      & \colhead{(deg)}            &                          
}
\startdata
2,454,957.18651    & 313.13               & 24.59                      & 0.925                    \\
2,454,964.49575    & 316.49               & 25.27                      & 0.911                    \\
2,454,971.80511    & 312.30               & 25.30                      & 0.921                    \\
2,454,979.11477    & 312.53               & 25.15                      & 0.906                    \\
2,454,986.42442    & 323.93               & 24.94                      & 0.920                    \\
2,454,993.73370    & 325.58               & 25.43                      & 0.906                    \\
2,455,001.04254    & 330.65               & 25.50                      & 0.872                    \\
2,455,008.35254    & 326.74               & 26.26                      & 0.923                    \\
2,455,015.66214    & 331.12               & 25.36                      & 0.935                    \\
2,455,022.97150    & 317.34               & 25.47                      & 0.917                    \\
\enddata
\tablecomments{This table is available in its entirety in machine-readable form. A portion is shown here for guidance regarding its form and content. }
\end{deluxetable}

\begin{deluxetable}{lrccccc}
\tabletypesize{\small}
\tablewidth{0pt}
\tablecaption{Multiple Frequency Analysis of V404 Lyr$\rm ^a$ }
\tablehead{
             & \colhead{Frequency}    & \colhead{Amplitude} & \colhead{Phase} & \colhead{S/N}  & \colhead{Remark$\rm ^b$}        \\
             & \colhead{(day$^{-1}$)} & \colhead{(mmag)}    & \colhead{(rad)} &                &
}                                                                                                            
\startdata                                                                                                   
$f_{1}$      &  1.97460$\pm$0.00001   & 6.34$\pm$0.22       & 1.88$\pm$0.10   & 73.19          &                                 \\
$f_{2}$      &  2.11165$\pm$0.00001   & 4.30$\pm$0.16       & 0.20$\pm$0.11   & 49.78          &                                 \\
$f_{3}$      &  2.08422$\pm$0.00001   & 3.87$\pm$0.17       & 2.51$\pm$0.13   & 44.90          &                                 \\
$f_{4}$      &  1.89372$\pm$0.00001   & 3.36$\pm$0.15       & 0.34$\pm$0.13   & 38.58          &                                 \\
$f_{5}$      &  2.03561$\pm$0.00001   & 3.36$\pm$0.17       & 0.45$\pm$0.15   & 38.97          &                                 \\
$f_{6}$      &  1.92343$\pm$0.00001   & 2.87$\pm$0.08       & 3.99$\pm$0.09   & 33.05          &                                 \\
$f_{7}$      &  5.47242$\pm$0.00001   & 2.54$\pm$0.20       & 1.03$\pm$0.23   & 58.67          & $4f_{\rm orb}$                  \\
$f_{8}$      &  1.12092$\pm$0.00001   & 2.20$\pm$0.24       & 2.76$\pm$0.32   & 22.12          &                                 \\
$f_{9}$      &  1.85447$\pm$0.00002   & 1.38$\pm$0.19       & 2.04$\pm$0.40   & 15.61          &                                 \\
$f_{10}$     &  3.49777$\pm$0.00002   & 0.94$\pm$0.15       & 2.81$\pm$0.46   & 12.68          & $4f_{\rm orb}-f_1$              \\
$f_{11}$     &  1.10448$\pm$0.00003   & 1.07$\pm$0.20       & 0.77$\pm$0.55   & 10.81          &                                 \\
$f_{12}$     &  0.60658$\pm$0.00002   & 0.95$\pm$0.14       & 2.03$\pm$0.42   &  9.19          & $f_1-f_{\rm orb}$               \\
$f_{13}$     &  3.36071$\pm$0.00002   & 0.90$\pm$0.15       & 4.54$\pm$0.47   & 11.90          & $4f_{\rm orb}-f_2$              \\
$f_{14}$     &  3.38817$\pm$0.00003   & 0.80$\pm$0.15       & 6.22$\pm$0.55   & 10.61          & $4f_{\rm orb}-f_3$              \\
$f_{15}$     &  2.74491$\pm$0.00003   & 0.78$\pm$0.16       & 3.25$\pm$0.60   &  9.17          &                                 \\
$f_{16}$     &  1.73291$\pm$0.00002   & 0.77$\pm$0.13       & 5.31$\pm$0.50   &  8.32          & $f_4+f_6-f_3$                   \\
$f_{17}$     &  3.43677$\pm$0.00002   & 0.75$\pm$0.13       & 4.81$\pm$0.51   & 10.03          & $4f_{\rm orb}-f_5$              \\
$f_{18}$     &  3.57865$\pm$0.00003   & 0.71$\pm$0.18       & 2.09$\pm$0.74   &  9.79          & $4f_{\rm orb}-f_4$              \\
$f_{19}$     &  1.12864$\pm$0.00003   & 0.67$\pm$0.16       & 1.52$\pm$0.68   &  6.76          &                                 \\
$f_{20}$     &  1.70321$\pm$0.00003   & 0.69$\pm$0.16       & 1.33$\pm$0.67   &  7.37          &                                 \\
$f_{21}$     &  2.73351$\pm$0.00003   & 0.64$\pm$0.16       & 4.03$\pm$0.71   &  7.50          &                                 \\
$f_{22}$     &  1.08650$\pm$0.00003   & 0.70$\pm$0.13       & 3.74$\pm$0.55   &  7.02          &                                 \\
$f_{23}$     &  3.54895$\pm$0.00004   & 0.66$\pm$0.18       & 3.45$\pm$0.81   &  9.08          & $4f_{\rm orb}-f_6$              \\
$f_{24}$     &  1.09571$\pm$0.00003   & 0.62$\pm$0.15       & 1.49$\pm$0.70   &  6.18          &                                 \\
$f_{25}$     &  2.73411$\pm$0.00003   & 0.61$\pm$0.14       & 0.63$\pm$0.69   &  7.11          &                                 \\
$f_{26}$     &  2.73052$\pm$0.00004   & 0.62$\pm$0.18       & 0.85$\pm$0.85   &  7.29          &                                 \\
$f_{27}$     &  1.86618$\pm$0.00004   & 0.52$\pm$0.14       & 0.15$\pm$0.80   &  5.88          & $f_3+f_4-f_2$                   \\
$f_{28}$     &  2.73164$\pm$0.00001   & 0.53$\pm$0.04       & 2.90$\pm$0.20   &  6.25          &                                 \\
$f_{29}$     & 10.94484$\pm$0.00004   & 0.55$\pm$0.15       & 0.05$\pm$0.79   & 25.67          & $8f_{\rm orb}$                  \\
$f_{30}$     &  2.74766$\pm$0.00004   & 0.54$\pm$0.17       & 4.68$\pm$0.93   &  6.31          &                                 \\
$f_{31}$     &  1.70811$\pm$0.00001   & 0.50$\pm$0.04       & 0.80$\pm$0.26   &  5.41          &                                 \\
$f_{32}$     &  4.71079$\pm$0.00001   & 0.98$\pm$0.08       & 2.85$\pm$0.24   & 19.49          & $2f_{\rm orb}+f_1$              \\
$f_{33}$     &  9.57666$\pm$0.00005   & 0.49$\pm$0.19       & 3.67$\pm$1.12   & 21.19          & $7f_{\rm orb}$                  \\
$f_{34}$     &  1.99260$\pm$0.00004   & 0.39$\pm$0.12       & 2.44$\pm$0.90   &  4.49          & $3f_{\rm orb}-f_2$              \\
$f_{35}$     &  2.72538$\pm$0.00005   & 0.47$\pm$0.16       & 1.28$\pm$1.01   &  5.55          &                                 \\
$f_{36}$     &  1.84243$\pm$0.00005   & 0.44$\pm$0.16       & 0.05$\pm$1.09   &  4.97          & $f_4+f_6-f_1$                   \\
$f_{37}$     &  1.07712$\pm$0.00005   & 0.46$\pm$0.18       & 1.12$\pm$1.12   &  4.61          &                                 \\
$f_{38}$     &  1.06724$\pm$0.00005   & 0.45$\pm$0.16       & 1.19$\pm$1.07   &  4.48          &                                 \\
$f_{39}$     &  1.21302$\pm$0.00005   & 0.42$\pm$0.15       & 4.00$\pm$1.05   &  4.33          & $2f_1-2f_{\rm orb}$             \\
$f_{40}$     &  2.02007$\pm$0.00005   & 0.41$\pm$0.15       & 2.27$\pm$1.04   &  4.78          & $3f_{\rm orb}-f_3$              \\
$f_{41}$     &  2.73900$\pm$0.00005   & 0.41$\pm$0.16       & 5.54$\pm$1.16   &  4.81          &                                 \\
$f_{42}$     &  1.67411$\pm$0.00005   & 0.41$\pm$0.15       & 2.80$\pm$1.09   &  4.33          &                                 \\
$f_{43}$     &  1.95660$\pm$0.00005   & 0.40$\pm$0.16       & 3.70$\pm$1.18   &  4.62          &                                 \\
$f_{44}$     &  1.72604$\pm$0.00005   & 0.39$\pm$0.14       & 4.52$\pm$1.06   &  4.23          &                                 \\
$f_{45}$     &  1.99353$\pm$0.00006   & 0.41$\pm$0.18       & 3.80$\pm$1.27   &  4.71          & $3f_{\rm orb}-f_2$              \\
\enddata
\tablenotetext{a}{Uncertainties were calculated according to Kallinger et al. (2008). }
\tablenotetext{b}{Possible harmonic and combination frequencies. }
\end{deluxetable}

\begin{deluxetable}{lcccccccc}
\tablewidth{0pt} 
\tablecaption{Absolute Parameters of V404 Lyr }
\tablehead{
\colhead{Parameter}                      & \multicolumn{2}{c}{Lee et al. (2014)}       && \multicolumn{2}{c}{This Paper}              \\ [1.0mm] \cline{2-3} \cline{5-6} \\[-2.0ex]
                                         & \colhead{Primary} & \colhead{Secondary}     && \colhead{Primary} & \colhead{Secondary}         
}                                                                                                                                     
\startdata
$M$ ($M_\odot$)                          & 1.35              & 0.52                    && 2.168$\pm$0.057    & 1.417$\pm$0.036        \\
$R$ ($R_\odot$)                          & 1.76              & 1.26                    && 1.909$\pm$0.023    & 1.791$\pm$0.021        \\
$\log$ $g$ (cgs)                         & 4.08              & 3.95                    && 4.212$\pm$0.015    & 4.083$\pm$0.015        \\
$\rho$ ($\rho_\odot$)                    & 0.25              & 0.26                    && 0.312$\pm$0.015    & 0.247$\pm$0.011        \\
$T$ (K)                                  & 6555              & 5362                    && 7330$\pm$150       & 5618$\pm$82            \\
$L$ ($L_\odot$)                          & 5.12              & 1.17                    && 9.43$\pm$0.80      & 2.86$\pm$0.18          \\
$M_{\rm bol}$ (mag)                      & 2.96              & 4.56                    && 2.294$\pm$0.093    & 3.588$\pm$0.069        \\
BC (mag)                                 & $+$0.01           & $-$0.18                 && 0.035$\pm$0.001    & $-$0.111$\pm$0.018     \\
$M_{\rm V}$ (mag)                        & 2.95              & 4.74                    && 2.259$\pm$0.093    & 3.700$\pm$0.071        \\
Distance (pc)                            & \multicolumn{2}{c}{465}                     && \multicolumn{2}{c}{671$\pm$41}              \\
\enddata
\end{deluxetable}

\end{document}